\documentclass[epj]{svjour}

\usepackage{graphics}
\graphicspath{{figures/}}

\usepackage{mathtools,slashed,amssymb,amsmath, bbm, caption,flushend, soul}

\usepackage[colorlinks,citecolor=blue,urlcolor=blue,linkcolor=blue]{hyperref}
\RequirePackage[T1]{fontenc}
\RequirePackage{mathptmx} 
\usepackage{mathtools,slashed}
\RequirePackage{graphicx}
\usepackage{adjustbox}
\usepackage{multirow}
\graphicspath{{images/}}

\RequirePackage{flushend}

\RequirePackage[numbers,sort&compress]{natbib}
\RequirePackage[colorlinks,citecolor=blue,urlcolor=blue,linkcolor=blue]{hyperref}

\RequirePackage{amsmath, amssymb, bbm, caption, mathtools}

\RequirePackage{xspace}

\newcommand\T{\rule{0pt}{2.6ex}}              \newcommand\B{\rule[-1.2ex]{0pt}{0pt}}   
\newcommand*{\TeV}{\ensuremath{\text{Te\kern -0.1em V}}\xspace}
\newcommand*{\GeV}{\ensuremath{\text{Ge\kern -0.1em V}}\xspace}
\newcommand{\lhcenergyComb}{\ensuremath{ 5.02\,\TeV}}
\newcommand{\cmenergyComb}{\ensuremath{\lhcenergyComb}\xspace}
\newcommand{\invnb}{\ensuremath{\rm nb^{-1}}\xspace}

\RequirePackage[version=3]{mhchem}
\usepackage{nag}
\newcommand{\PbPb}{\ce{Pb}+\ce{Pb}\xspace}
\newcommand{\pp}{\ensuremath{pp}\xspace}
\newcommand{\sqrtNN}{\ensuremath{\sqrt{\smash[b]{s_{_{\mathrm{NN}}}}}}\xspace}

\newcommand{\mygg}{\ensuremath{\gamma\gamma}\xspace}
\newcommand{\myprocess}{\ensuremath{\PbPb\,(\mygg)\to \textrm{Pb}^{(\ast)}\textrm{+}\textrm{Pb}^{(\ast)}\,\mygg
}}
\newcommand{\Et}{\mbox{$E_{\mathrm{T}}$}\xspace}
\newcommand{\Minvgg}{\mbox{$m_{\mygg}$}\xspace}
\newcommand{\pT}{\mbox{$p_{\mathrm{T}}$}\xspace}
\newcommand{\ptgg}{\mbox{$p_{\mathrm{T}}^{\mygg}$}\xspace}
\newcommand{\gggg}{\mbox{$\mygg\to \mygg$}\xspace}
\newcommand{\ggee}{\mbox{$\mygg\to \ee$}\xspace}
\newcommand{\ee}{\mbox{$e^+e^-$}\xspace}
\newcommand{\axion}{\mbox{$a$}\xspace}

\newcommand{\Aco}{\mbox{$A_{\phi}$}\xspace}

\newcommand{\sigmafid}{\ensuremath{\sigma_{\text{fid.}}}\xspace}
\newcommand{\sigmaraw}{\ensuremath{\sigma^{\text{fid.}}_{\text{raw}}}\xspace}
\newcommand{\sigmascaled}{\ensuremath{\sigma^{\text{fid.}}_{\text{cor.}}}\xspace}
\newcommand{\sigmameas}{\ensuremath{\sigma_{\text{meas.}}^{\text{fid.}}}\xspace}
\newcommand{\sigmatheo}{\ensuremath{\sigma_{\text{theo.}}^{\text{fid.}}}\xspace}

\newcommand{\finalResult}{\ensuremath{\sigmameas &= 115\pm 15\;(\text{stat.})\pm 11\;(\text{syst.})\pm 3\;(\text{lumi.})\pm 3\;(\text{theo.})\,\,\text{nb} \\ &= 115\pm 19\,\,\text{nb}}}
\newcommand{\finalResultTotUnct}{\ensuremath{115\pm 19\,\,\text{nb}}}
\newcommand{\finalResultCentVal}{\ensuremath{115\,\text{nb}}}
\newcommand{\finalResultAbsPrecision}{\ensuremath{19\,\,\text{nb}}}
\newcommand{\finalResultRelPrecision}{\ensuremath{17\%}}

\newcommand{\blueOverallCorrProb}{\ensuremath{7.6}\xspace}

\begin{document}\sloppy
\title{Light-by-light scattering cross-section measurements at LHC}

\author{
{\large G.~K. Krintiras\href{https://orcid.org/0000-0002-0380-7577}{\includegraphics[scale=0.15]{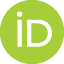}}}\inst{1}\thanks{ gkrintir@cern.ch}
\and {\large I. Grabowska-Bold\href{https://orcid.org/0000-0001-9159-1210}{\includegraphics[scale=0.15]{figures/or.png}}}\inst{2}\thanks{Iwona.Grabowska@cern.ch}
\and {\large M. K{\l}usek-Gawenda\inst{3}}\thanks{ mariola.klusek@ifj.edu.pl}
\and {\large \'{E}.~Chapon}\inst{4}\thanks{emilien.chapon@cern.ch} 
\and {\large R. Chudasama}\inst{5}\thanks{ruchi.chudasama@cern.ch} 
\and {\large R. Granier de Cassagnac}\inst{6}\thanks{raphael@cern.ch} 
}

\institute{Department of Physics and Astronomy, The University of Kansas, Malott Hall, 1251 Wescoe Hall Dr., Lawrence, KS 66045 \label{1}
\and Department of Physics and Applied Computer Science, AGH University of Science and Technology, Al. Mickiewicza 30, 30-059 Krak\'ow \label{2}
\and Institute of Nuclear Physics Polish Academy of Sciences, Radzikowskiego 152, PL-31-342 Krak\'ow, Poland \label{3}
\and IRFU, CEA, Universit\'{e} Paris-Saclay, Gif-sur-Yvette, France\label{4}
\and Department of High Energy Physics, Tata Institute of Fundamental Research, Mumbai 400005, India\label{5}
\and Laboratoire Leprince-Ringuet, Ecole polytechnique, CNRS/IN2P3, Palaiseau, France\label{6}
}

\abstract{
 This note represents an attempt to gather the input related to light-by-light scattering ($\mygg$) cross-section measurements at LHC with the aim of checking the consistency with different standard model predictions. For the first time, we also consider the contribution from the $\eta_b(1S)$ meson production to the diphoton invariant mass distribution, by calculating its inclusive photoproduction cross-section. Using a simplified set of assumptions, we find a result of $\finalResultTotUnct$, consistent with standard model predictions within two standard deviations. Although an improved determination of the integrated fiducial $\myprocess$ cross-section by approximately 10\% could be potentially achieved relative to current measurements, further improvements are expected with the inclusion of existing or forthcoming LHC nuclear data.  
}
\maketitle

\section{Introduction}
\label{sec:intro}

\begin{figure}[h]
\centering
\includegraphics[width=0.45\textwidth]{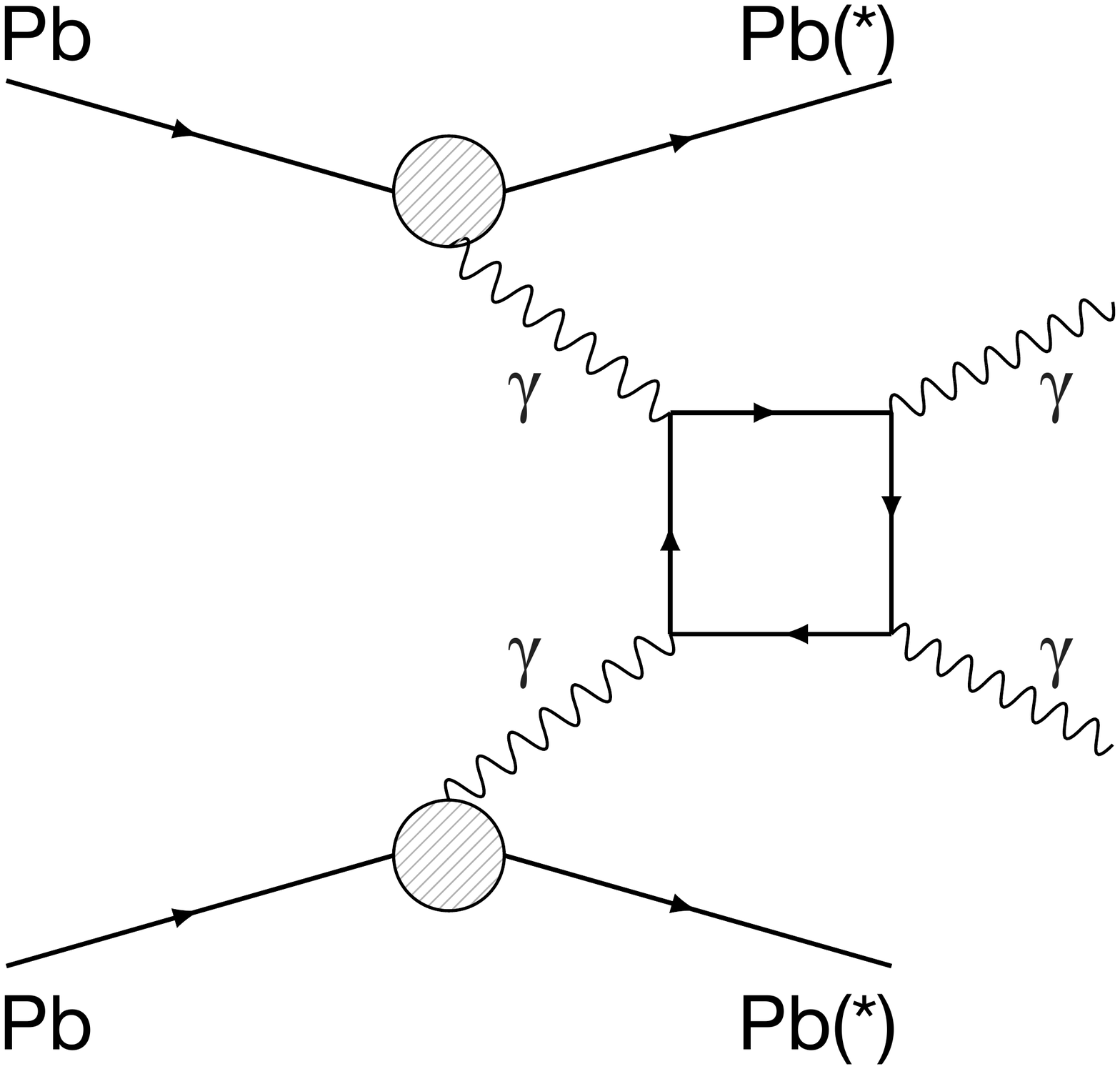}
\caption{\label{fig:UPC}The lowest order QED diagrams of SM LbyL scattering production in \PbPb\ UPC.
A potential EM excitation of the outgoing Pb ions is denoted by $(^{\ast})$. Plot adapted from Ref.~\cite{Aad:2020cje}.}
\end{figure}

Light-by-light~(LbyL) scattering, $\mygg \to \mygg$, is a rare standard model~(SM) process that can occur at the lowest order in quantum electrodynamics~(QED) via virtual one-loop box diagrams~(Figure~\ref{fig:UPC}). Experimentally, we can directly study LbyL interactions at relativistic heavy-ion collisions with large impact parameters, i.e., larger than twice the radius of the ions, as the strong interaction in these ultra-peripheral collision~(UPC) events is diminished. The electromagnetic (EM) fields produced by the colliding nuclei can be treated as a beam of quasi-real photons, resulting in the enhancement of the cross-section for the reaction $\textrm{A+A}\,(\mygg) \to \textrm{A}^{(\ast)}\textrm{+}\textrm{A}^{(\ast)}\,\mygg$ relative to proton-proton (\pp) collisions by a factor of $Z^4$~\cite{Baltz:2007kq, Enterria:2013yra, Klein:2020fmr}. The signature of the LbyL final state is the ``exclusive'' production of two photons,  where the incoming ions survive the EM interaction: two photons with low transverse energy are expected and no further activity in the central detector surrounding the interaction region. The LbyL process has been also proposed as a sensitive channel to study physics beyond the SM~\cite{Fichet:2014uka,Knapen:2016moh,Ellis:2017edi,Kostelecky:2018yfa,Inan:2019ugz,Horvat:2020ycy}. In particular, the enhancement of the diphoton  rates can be induced by new neutral particles, such as axion-like particles~(ALP), contributing in the form of narrow diphoton resonances.

The first direct evidence of the LbyL process at the LHC was established by the  ATLAS~\cite{Aaboud:2017bwk} and CMS~\cite{Sirunyan:2018fhl} Collaborations, exploiting the large photon fluxes available in UPC lead-lead (\PbPb) collisions at the LHC. The evidence was obtained from \PbPb data recorded in 2015 at a per nucleon center-of-mass energy of \sqrtNN~=~\cmenergyComb with integrated luminosities of 0.48~\invnb~(ATLAS) and 0.39~\invnb~(CMS). The CMS Collaboration also set upper limits on the cross-section for ALP production, $\mygg \to \axion \to \mygg$, over the ALP invariant mass range of 5--90~\GeV, whereas previous ATLAS searches involving ALP decays to photons were based on \pp collision data~\cite{ATLAS:2015rsn,ATLAS:2018dfo}. Exploiting a data sample of \PbPb\ collisions collected in 2018 at the same center-of-mass energy with an integrated luminosity of 1.73~\invnb, the ATLAS Collaboration observed the LbyL scattering process with a significancee larger than five standard deviations~\cite{Aad:2019ock}. An updated measurement was performed in Ref.~\cite{Aad:2020cje}, using a combination of \PbPb collision data recorded in 2015 and 2018 by the ATLAS experiment, and corresponding to an integrated luminosity of 2.2~$\textrm{nb}^{-1}$.
  
This note represents an attempt to average the integrated fiducial cross-sections for the \myprocess\ production at \sqrtNN~=~\cmenergyComb\ using the available LHC data sets. Section~\ref{sec:atlascmsxs} gives a brief description of the input cross-section measurements, while further details are given in~\ref{app:allUncs}. The theoretical cross-section calculations are described in Section~\ref{sec:theory}, including the $\eta_b(1S)$ meson photoproduction and compared to the $\gggg$ invariant mass distributions, typically used to set limits on ALP production. Section~\ref{sec:xscomb} presents the averaged cross-section, using a set of simplified assumptions given in~\ref{app:systcat}. In Section~\ref{sec:sum}, the results are summarized.

 \section{Input light-by-light cross-section measurements}
\label{sec:atlascmsxs}

In all analyses, candidate diphoton events were recorded using dedicated triggers aiming at low photon transverse energy, \Et,
for events with moderate activity in the calorimeter but little additional activity in the entire detectors to suppress nonexclusive background.
In the offline ATLAS (CMS) analysis used as input measurements, events were selected with exactly two photons, each with $\Et>2.5$ ($>$2.0)~\GeV\ and pseudorapidity $|\eta|<2.4$, that satisfied further selection requirements described in Refs.~\cite{Aad:2020cje,Sirunyan:2018fhl}. To suppress nonexclusive background, characterized by a final state with larger transverse momenta \pT
and diphoton acoplanarities, $\Aco = (1-|\Delta\phi^{\mygg}|/\pi)$, the \pT and $\Aco$ 
were required to be $\ptgg<1$~\GeV and $\Aco<0.01$, respectively, and the invariant mass of the pair to be $m_{\mygg}>5$~\GeV.
The two dominant exclusive background sources potentially remaining in the LbyL scattering signal region were
exclusive dielectron pairs from the reaction $\PbPb\,(\mygg)\to \textrm{Pb}^{(\ast)}\textrm{+}\textrm{Pb}^{(\ast)}\,e^+e^-$ and central exclusive production of the diphoton system $gg \to \mygg$. Their contribution was evaluated and subtracted. 

The cross-section for the \myprocess\ process was measured in a fiducial phase space, defined by the above requirements on the diphoton final state, and reflecting the selection at the reconstruction level. A summary of the available measurements along with their total uncertainty, evaluated as the quadratic sum of the individual sources, is presented in Table~\ref{tab:xsec_meas}. A series of analysis improvements were introduced in Ref.~\cite{Aad:2020cje}, leading to a broader kinematic range in diphoton invariant mass ($\Minvgg>5$ vs. $>6$~\GeV) and single-photon transverse energy ($\Et>2.5$ vs. $>3$~\GeV) relative to Refs.~\cite{Aaboud:2017bwk,Aad:2019ock}.  Details are given in~\ref{app:allUncs} for the input cross-sections and marked with $^\dagger$ in Table~\ref{tab:xsec_meas}. 

\begin{table*}[!htbp]
  \centering
  \caption{Summary of the fiducial LbyL cross-section measurements  at
    \cmenergyComb performed (``$\sigmaraw$'') by the ATLAS and CMS  Collaborations. When applicable, they are further scaled by correction factors (``$\sigmascaled$'') to account for differences in the definition of phase space regions, as described in Section~\ref{sec:theory}. Total uncertainties are shown.
    The symbol ``---'' means that no corresponding cross-section measurement currently exists.
    The cross-sections marked with $^\dagger$ are those used as input to the extraction of the averaged value of the \myprocess\ process.}
  \begin{adjustbox}{max width=1.0\textwidth}{}
    \begin{tabular}{l|l|r@{\hskip 0cm}lr@{\hskip 0cm}l|r@{\hskip 0cm}lr@{\hskip 0cm}l}
      \hline
      \hline
      \multicolumn{2}{c|}{} & \multicolumn{4}{c|}{ATLAS} & \multicolumn{4}{c}{CMS} \\
      \hline
      \sqrtNN & Year (Lumi.\ [$\invnb$]) & \multicolumn{2}{c}{\sigmaraw\,[nb]} & \multicolumn{2}{c}{\sigmascaled\,[nb]} & \multicolumn{2}{c}{\sigmaraw\,[nb]} & \multicolumn{2}{c}{\sigmascaled\,[nb]} \\
      \hline
                            & 2015 (0.39--0.48) & $70$&$~\pm~29$~\cite{Aaboud:2017bwk} & $108$&$~\pm~45$ & $120$&$~\pm~55$~\cite{Sirunyan:2018fhl} & $91$&$~\pm~42$${}^{\dagger}$\\ 
      \cline{2-10}
      \T\B
      %\hline
             \cmenergyComb                & 2018 (1.73) & $78$&$~\pm~15$~\cite{Aad:2019ock} & $120$&$~\pm~23$ & \multicolumn{2}{c}{---} & ---  \\ 
      \cline{2-10}
      \T\B
       & 2015$+$2018 (2.2) & $120$&$~\pm~22$~\cite{Aad:2020cje} & $120$&$~\pm~22$${}^{\dagger}$& \multicolumn{2}{c}{---} & --- \\
      \hline
      \hline
    \end{tabular}
    \renewcommand{\arraystretch}{1}
    \end{adjustbox}
  \label{tab:xsec_meas}
\end{table*}

\section{Theoretical LbyL cross-section calculations}
\label{sec:theory}

The theoretical predictions for the \myprocess\ production cross-section at \sqrtNN~=~\cmenergyComb\ are calculated numerically at leading order~(LO). More specifically, the SuperChic~v3.0~\cite{Harland-Lang:2018iur} Monte Carlo (MC) framework takes into account box diagrams with leptons and quarks (such as the diagram in the panel of Figure~\ref{fig:UPC}), and $W^{\pm}$ bosons, while including interference effects.
The $W^\pm$ contribution is only important for diphoton masses $\Minvgg > 2 m_{W}$. 

An alternative LbyL numerical calculation is performed in Ref.~\cite{Klusek-Gawenda:2016euz}, with the main difference relative to SuperChic being in the implementation of the nonhadronic-overlap condition of the Pb ions. In SuperChic, the probability for exclusive \mygg interactions turns on smoothly for \PbPb impact parameters in the range of 15--20~fm and it is unity for larger values, while the alternative prediction fully suppresses these interactions for impact parameters below 14~fm when two nuclei overlap during the collision. This different approach leads to a fiducial cross-section for LbyL scattering about 2--4\% larger in the alternative calculation than in the prediction from SuperChic. 

The theoretical uncertainty in these two calculations is primarily due to limited knowledge of the nuclear (EM) form-factors and the related initial photon fluxes, with no dependence on the diphoton mass for $\Minvgg<100$~\GeV. In both cases, next-to-leading-order (NLO) QCD and QED corrections~\cite{Bern:2001dg, Klusek-Gawenda:2016nuo} increase the cross-section by a few percents, and are taken into account in the quoted total uncertainties of 10\%. Yet other alternative calculations~\cite{Enterria:2013yra} based on the equivalent photon approximation, and using the MadGraph5 MC framework with the LO expression for the $\mygg \to \mygg$ cross-section for all quark and lepton loops, agree to the level of the quoted total uncertainties. 

The theoretical calculations are performed at a fixed center-of-mass energy while the energy of the LHC beam is measured with a relative uncertainty of 0.1\%~\cite{Todesco:2017nnk}, hence bearing negligibly small impact on LbyL cross-sections.

A summary of theoretical cross-section predictions with their uncertainties is shown in Table~\ref{tab:xs}, separately for the phase space regions defined in Section~\ref{sec:atlascmsxs}. Based on these numbers, the ratios of $77/101\approx{0.76}$ and $77/50\approx{1.54}$ are used as  correction factors accounting for differences in the definition of phase space regions between experiments and among analyses, respectively. The theoretical cross-section that is used as reference is marked with $^\dagger$.

\begin{table*}[!htb]
  \centering
  \caption{Predicted cross-sections for LbyL scattering at \cmenergyComb. Uncertainties take into account derivations from alternative approaches.
    The cross-section marked with $^\dagger$ is used as reference.}
  \begin{adjustbox}{max width=1.0\textwidth}{}
    \begin{tabular}{l|l|l|l|l}
      \hline
      \hline
      \T\B
      \sqrtNN & Process & Accuracy & $\sigmatheo{}$\,[nb] & Phase space region\\
      \hline
      \T\B
        &  & LO & $101 \pm 10$~\cite{Harland-Lang:2018iur} & $\Et>2.0$~\GeV, $|\eta|<2.4$, $\Minvgg>5$~\GeV, $\pT^{\mygg}<1$~\GeV, $\Aco<0.01$\\
      \cline{3-4}
      \T\B
       &  & LO & $103 \pm 10$~\cite{Klusek-Gawenda:2016euz} & $\Et>2.0$~\GeV, $|\eta|<2.4$, $\Minvgg>5$~\GeV, $\pT^{\mygg}<1$~\GeV, $\Aco<0.01$\\
      \cline{3-4}
      \T\B
      \cmenergyComb & \myprocess& LO & $77 \pm 8{}^{\dagger}$~\cite{Harland-Lang:2018iur} & $\Et>2.5$~\GeV, $|\eta|<2.4$, $\Minvgg>5$~\GeV, $\pT^{\mygg}<1$~\GeV, $\Aco<0.01$\\
      \cline{3-4}
      \T\B
        & & LO & $80 \pm 8{}$~\cite{Klusek-Gawenda:2016euz}& $\Et>2.5$~\GeV, $|\eta|<2.4$, $\Minvgg>5$~\GeV, $\pT^{\mygg}<1$~\GeV, $\Aco<0.01$\\
        \cline{3-4}
      \T\B
                 & & LO & $50 \pm 5$~\cite{Harland-Lang:2018iur} & $\Et>3.0$~\GeV, $|\eta|<2.4$, $\Minvgg>6$~\GeV, $\pT^{\mygg}<1$~\GeV, $\Aco<0.01$ \\
      \cline{3-4}
      \T\B
                 & & LO & $51 \pm 5$~\cite{Klusek-Gawenda:2016euz} & $\Et>3.0$~\GeV, $|\eta|<2.4$, $\Minvgg>6$~\GeV, $\pT^{\mygg}<1$~\GeV, $\Aco<0.01$ \\
      \hline
      \hline
    \end{tabular}
    \end{adjustbox}{}
  \label{tab:xs}
\end{table*}

\begin{figure*}[!h!tbp]
  \begin{center}
    \includegraphics[width=0.5\textwidth]{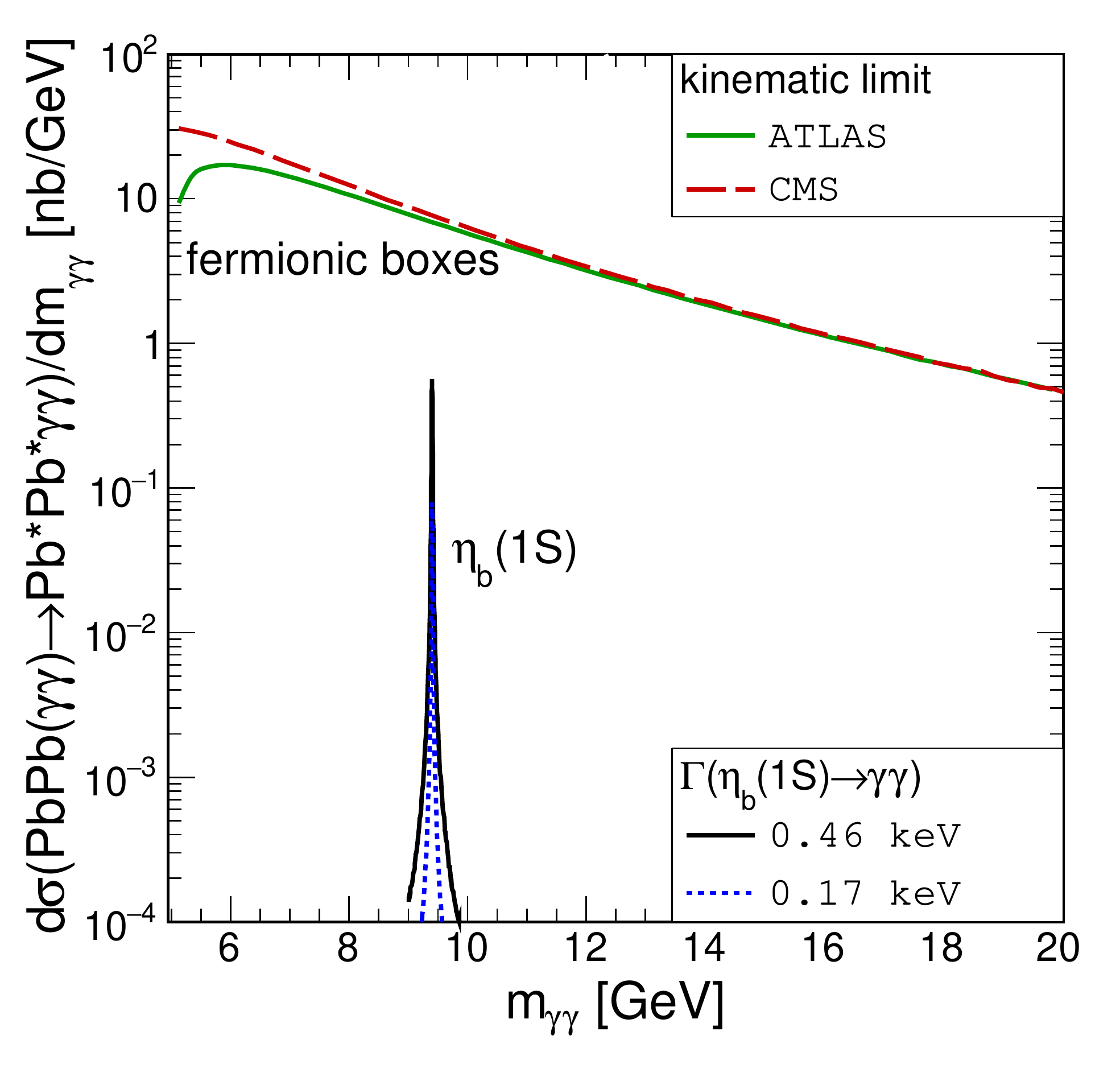}
    \caption{Differential cross-sections for the signal (contribution from fermionic boxes~\cite{Klusek-Gawenda:2016euz}) \myprocess\ and the intermediate $\mygg\rightarrow \eta_b(1S)\to \mygg$ background production process,  as a function of the diphoton invariant mass. For the signal, the ATLAS and CMS kinematic requirements from Refs.~\cite{Aad:2020cje} and~\cite{Sirunyan:2018fhl}, respectively, are adopted.  For the background $\eta_b(1S)$ process, the decay to a diphoton system is shown for the maximum and minimum values of diphoton decay rates of 0.46~\cite{Gupta:1996ak} and 0.17~keV~\cite{Ackleh:1991dy}, respectively. 
    }
    \label{fig:etaB}
\end{center}
\end{figure*}

Differential distributions involving kinematic variables of the final-state photons, in particular the invariant mass of the diphoton system, characterize the energy of the process (typically used to set limits on ALP production). Figure~\ref{fig:etaB} shows a comparison of the theoretical calculations~\cite{Klusek-Gawenda:2016euz} of differential cross-sections as a function of the invariant mass of the diphoton system for \gggg\ scattering and $\eta_b(1S)$ meson production with decays into two photons in the final state. The importance of $\eta_b(1S)$ meson is studied for the \myprocess\ process for the first time. According to the Particle Data Group~\cite{ParticleDataGroup:2020ssz}, the $\eta_b(1S)$ meson has well-defined mass and width, but the $\mygg$ decay width is still being studied theoretically (see e.g., Ref.~\cite{Ebert:2003mu} and references therein) and has not been experimentally verified.
Theoretical uncertainties originate from different values of the $\eta_b(1S)\to\mygg$ decay width. The maximum and minimum values of diphoton decay rates are 0.46~\cite{Gupta:1996ak} and 0.17~keV~\cite{Ackleh:1991dy}, respectively.
Although the height of the resonance peak is $1$~nb, this contribution is therefore found to be insignificant in the context of the LbyL measurement because the experimental \Minvgg\ bins are typically very wide. The total nuclear cross section for the $\mygg \to \eta_b(1S) \to \mygg$ resonance scattering depending on the decay width is
$\sigma_{\text{theo.}}^{\text{tot.}}\left(\ensuremath{\PbPb\,(\mygg\to\eta_b(1S))\to \textrm{Pb}^{(\ast)}\textrm{+}\textrm{Pb}^{(\ast)}\,\mygg}\right)=\left( 0.19 - 1.41 \right)10^{-2}$~nb. It is also worth highlighting the influence of the lower $\Et$ requirement on diphoton invariant mass, resulting in increased acceptance at low ($\lesssim 6\,\GeV$) \Minvgg\ values. The difference in the imposed requirements on transverse energy, i.e., $\Et>2.5$ and $>2$~\GeV for ATLAS and CMS, respectively, does not affect the $\eta_b(1S)$ distribution because of the $2\Et < m_{\eta_b(1S)}$ condition.

\section{Averaged cross-section measurement}
\label{sec:xscomb}
The cross-section measurements, described in
Section~\ref{sec:atlascmsxs} and denoted by $^\dagger$ in Table~\ref{tab:xsec_meas}, are used as input to an averaged cross-section. We use the best linear unbiased estimator (BLUE) method~\cite{Lyons:1988rp, Valassi:2003mu,Nisius:2014wua}. More specifically, the BLUE software~v2.4.0, as implemented within the ROOT analysis framework~\cite{Nisius:2020jmf}, is used. 
Systematic uncertainties are categorized, and simplified correlation assumptions are employed according to~\ref{app:systcat}.

The averaged cross-section measurement
at \cmenergyComb, after one iteration, is 
\begin{align*}
\finalResult,
\end{align*}
with a relative uncertainty of $\finalResultRelPrecision$. The overall correlation between the input measurements is \blueOverallCorrProb\%, hence are dominantly uncorrelated. The statistical uncertainty is still found to be the dominant overall uncertainty. The contribution from each uncertainty category to the total uncertainty in the averaged cross-section measurement is given in~\ref{app:systcat}.

\begin{figure*}[!h!tbp]
  \begin{center}
    \includegraphics[width=0.82\textwidth]{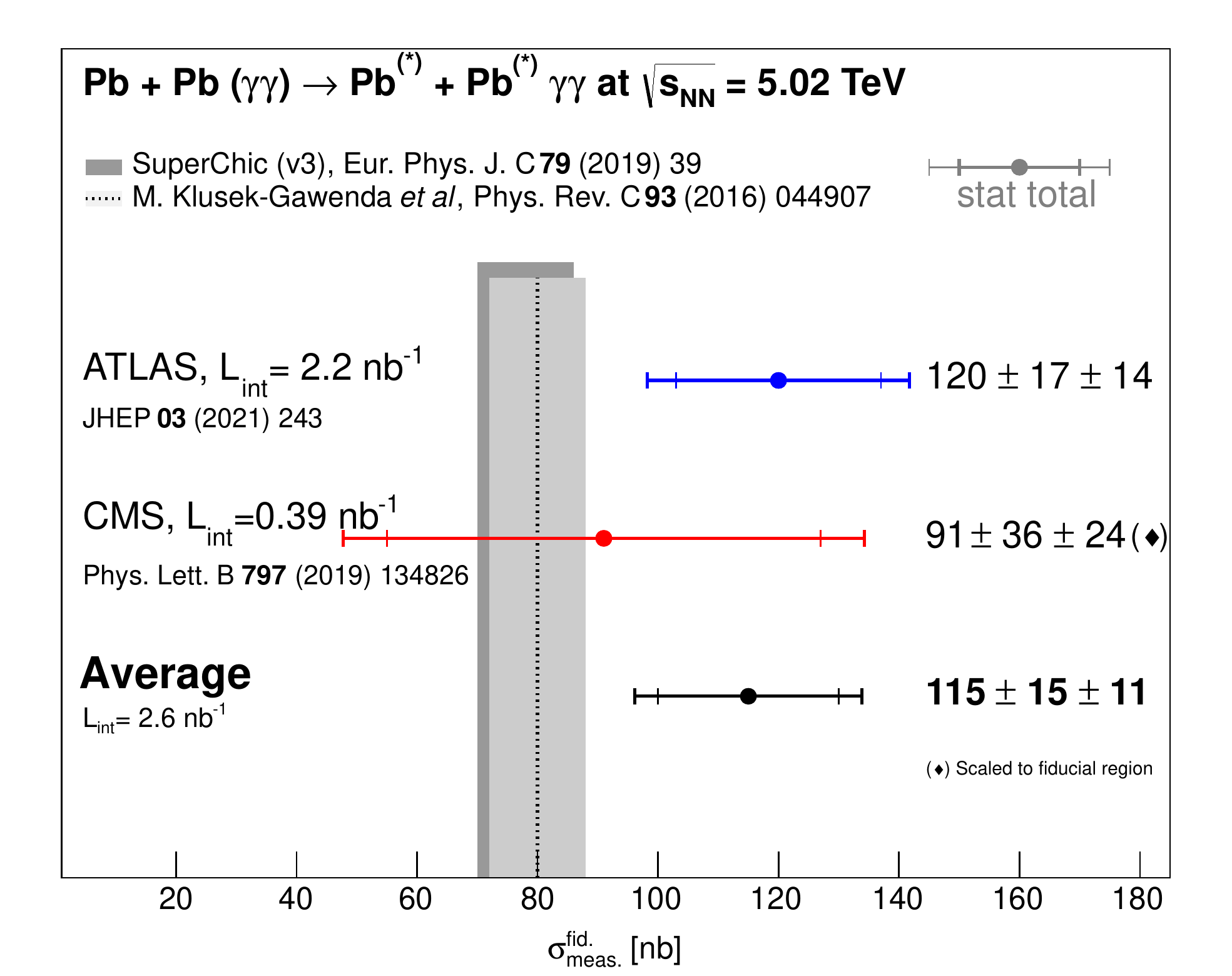}
    \caption{The averaged $\myprocess$ cross-section value along with the individual cross-section measurements at \cmenergyComb\ from ATLAS and CMS.
    The theoretical predictions~\cite{Harland-Lang:2018iur,Klusek-Gawenda:2016euz} are computed at LO accuracy. The $\sigmatheo$ uncertainties used to compute $\myprocess$ are described in the text.
    }
    \label{fig:result}
\end{center}
\end{figure*}

Figure~\ref{fig:result} shows a summary of the $\myprocess$ measurements at \cmenergyComb\, and their comparison to the theory predictions. The averaged cross-section is consistent within about two standard deviations with the SM predictions. 

\section{Summary}
\label{sec:sum}
This note represents an attempt to average existing measurements of the light-by-light (\gggg) scattering process from ultra-peripheral lead-lead collisions at \sqrtNN~=~\cmenergyComb\ at the LHC. Using a simplified set of assumptions, the integrated fiducial cross-section of the $\myprocess$ process is found to be \begin{align*} \finalResult.\end{align*}
The averaged cross-section is consistent within two standard deviations with standard model predictions, although the precision is still limited by the statistical uncertainty.  To check the possibility of unaccounted background sources, and for the first time, we calculate the potential contribution of the $\eta_b(1S)$ meson production to the measured \gggg invariant mass distributions, typically used to search for axion-like particles and set new exclusion limits. 

This result paves the way for combining existing or forthcoming measurements using LHC heavy-ion collisions and provides, within the studied phase space region, an additional experimental input to the comparison with state-of-the-art predictions from quantum electrodynamics.

\section{Acknowledgements}

This article is part of a project that has received funding from the European Union’s Horizon 2020 research and innovation program under the grant agreement No 824093, better known as STRONG–2020. GKK is supported by the Office of Nuclear Physics in the Department of Energy (DOE NP) under grant number \href{https://pamspublic.science.energy.gov/WebPAMSExternal/Interface/Common/ViewPublicAbstract.aspx?rv=00d4fe0f-48a0-4d4a-baf1-c70867d9e499&rtc=24&PRoleId=10}{DE-FG02-96ER40981}. IGB is supported by the National Science Centre of Poland under grant number UMO-2020/37/B/ST2/01043 and by PL-GRID infrastructure.

\bibliography{Letter}

\begin{thebibliography}{30}
\providecommand{\natexlab}[1]{#1}
\providecommand{\url}[1]{\texttt{#1}}
\expandafter\ifx\csname urlstyle\endcsname\relax
  \providecommand{\doi}[1]{doi: #1}\else
  \providecommand{\doi}{doi: \begingroup \urlstyle{rm}\Url}\fi

\bibitem[Baltz et~al.(2008)Baltz, Baur, d'Enterria, Frankfurt, Gelis, Guzey,
  Hencken, Kharlov, Klasen, Klein, Nikulin, et~al.]{Baltz:2007kq}
A.~J. Baltz, G.~Baur, D.~d'Enterria, L.~Frankfurt, F.~Gelis, V.~Guzey,
  K.~Hencken, Yu. Kharlov, M.~Klasen, S.~R. Klein, V.~Nikulin, et~al.
\newblock {The physics of ultraperipheral collisions at the LHC}.
\newblock \emph{Phys. Rept.}, 458:\penalty0 1, 2008.
\newblock \doi{10.1016/j.physrep.2007.12.001}.

\bibitem[d'Enterria and da~Silveira(2013)]{Enterria:2013yra}
David d'Enterria and Gustavo~G. da~Silveira.
\newblock {Observing Light-by-Light Scattering at the Large Hadron Collider}.
\newblock \emph{Phys. Rev. Lett.}, 111:\penalty0 080405, 2013.
\newblock \doi{10.1103/PhysRevLett.111.080405}.
\newblock [\href{https://dx.doi.org/10.1103/PhysRevLett.116.129901}{Erratum:
  Phys. Rev. Lett. \textbf{116} (2016) 129901}].

\bibitem[Klein and Steinberg(2020)]{Klein:2020fmr}
Spencer~R. Klein and Peter Steinberg.
\newblock Photonuclear and two-photon interactions at high-energy nuclear
  colliders.
\newblock \emph{Annu. Rev. Nucl. Part. Sci.}, 70\penalty0 (1):\penalty0
  323--354, 2020.
\newblock \doi{10.1146/annurev-nucl-030320-033923}.

\bibitem[Fichet et~al.(2015)Fichet, von Gersdorff, Lenzi, Royon, and
  Saimpert]{Fichet:2014uka}
Sylvain Fichet, Gero von Gersdorff, Bruno Lenzi, Christophe Royon, and Matthias
  Saimpert.
\newblock {Light-by-light scattering with intact protons at the LHC: from
  standard model to new physics}.
\newblock \emph{JHEP}, 02:\penalty0 165, 2015.
\newblock \doi{10.1007/JHEP02(2015)165}.

\bibitem[Knapen et~al.(2017)Knapen, Lin, Lou, and Melia]{Knapen:2016moh}
Simon Knapen, Tongyan Lin, Hou~Keong Lou, and Tom Melia.
\newblock {Searching for Axionlike Particles with Ultraperipheral Heavy-Ion
  Collisions}.
\newblock \emph{Phys. Rev. Lett.}, 118\penalty0 (17):\penalty0 171801, 2017.
\newblock \doi{10.1103/PhysRevLett.118.171801}.

\bibitem[Ellis et~al.(2017)Ellis, Mavromatos, and You]{Ellis:2017edi}
John Ellis, Nick~E. Mavromatos, and Tevong You.
\newblock {Light-by-Light Scattering Constraint on Born-Infeld Theory}.
\newblock \emph{Phys. Rev. Lett.}, 118\penalty0 (26):\penalty0 261802, 2017.
\newblock \doi{10.1103/PhysRevLett.118.261802}.

\bibitem[Kostelecky and Li(2019)]{Kostelecky:2018yfa}
V.~Alan Kostelecky and Zonghao Li.
\newblock {Gauge field theories with Lorentz-violating operators of arbitrary
  dimension}.
\newblock \emph{Phys. Rev. D}, 99:\penalty0 056016, 2019.
\newblock \doi{10.1103/PhysRevD.99.056016}.

\bibitem[Inan and Kisselev(2019)]{Inan:2019ugz}
S.~C. Inan and A.~V. Kisselev.
\newblock {Probe of the Randall-Sundrum-like model with the small curvature via
  light-by-light scattering at the LHC}.
\newblock \emph{Phys. Rev. D}, 100\penalty0 (9):\penalty0 095004, 2019.
\newblock \doi{10.1103/PhysRevD.100.095004}.

\bibitem[Horvat et~al.(2020)Horvat, Latas, Trampeti\'c, and
  You]{Horvat:2020ycy}
Raul Horvat, Du\v~sko Latas, Josip Trampeti\'c, and Jiangyang You.
\newblock {Light-by-light scattering and spacetime noncommutativity}.
\newblock \emph{Phys. Rev. D}, 101\penalty0 (9):\penalty0 095035, 2020.
\newblock \doi{10.1103/PhysRevD.101.095035}.

\bibitem[{ATLAS Collaboration}(2021)]{Aad:2020cje}
{ATLAS Collaboration}.
\newblock {Measurement of light-by-light scattering and search for axion-like
  particles with 2.2 nb$^{-1}$ of Pb+Pb data with the ATLAS detector}.
\newblock \emph{JHEP}, 03:\penalty0 243, 2021.
\newblock \doi{10.1007/JHEP03(2021)243}.

\bibitem[{ATLAS Collaboration}(2017)]{Aaboud:2017bwk}
{ATLAS Collaboration}.
\newblock {Evidence for light-by-light scattering in heavy-ion collisions with
  the ATLAS detector at the LHC}.
\newblock \emph{Nature Phys.}, 13\penalty0 (9):\penalty0 852--858, 2017.
\newblock \doi{10.1038/nphys4208}.

\bibitem[{CMS Collaboration}(2019)]{Sirunyan:2018fhl}
{CMS Collaboration}.
\newblock {Evidence for light-by-light scattering and searches for axion-like
  particles in ultraperipheral PbPb collisions at $\sqrt{s_\mathrm{NN}} =$ 5.02
  TeV}.
\newblock \emph{Phys. Lett. B}, 797:\penalty0 134826, 2019.
\newblock \doi{10.1016/j.physletb.2019.134826}.

\bibitem[{ATLAS Collaboration}(2016)]{ATLAS:2015rsn}
{ATLAS Collaboration}.
\newblock {Search for new phenomena in events with at least three photons
  collected in $pp$ collisions at $\sqrt{s}$ = 8 TeV with the ATLAS detector}.
\newblock \emph{Eur. Phys. J. C}, 76\penalty0 (4):\penalty0 210, 2016.
\newblock \doi{10.1140/epjc/s10052-016-4034-8}.

\bibitem[{ATLAS Collaboration}(2019{\natexlab{a}})]{ATLAS:2018dfo}
{ATLAS Collaboration}.
\newblock {A search for pairs of highly collimated photon-jets in $pp$
  collisions at $\sqrt{s}$ = 13 TeV with the ATLAS detector}.
\newblock \emph{Phys. Rev. D}, 99\penalty0 (1):\penalty0 012008,
  2019{\natexlab{a}}.
\newblock \doi{10.1103/PhysRevD.99.012008}.

\bibitem[{ATLAS Collaboration}(2019{\natexlab{b}})]{Aad:2019ock}
{ATLAS Collaboration}.
\newblock {Observation of Light-by-Light Scattering in Ultraperipheral Pb+Pb
  Collisions with the ATLAS Detector}.
\newblock \emph{Phys. Rev. Lett.}, 123\penalty0 (5):\penalty0 052001,
  2019{\natexlab{b}}.
\newblock \doi{10.1103/PhysRevLett.123.052001}.

\bibitem[Harland-Lang et~al.(2019)Harland-Lang, Khoze, and
  Ryskin]{Harland-Lang:2018iur}
L.~A. Harland-Lang, V.~A. Khoze, and M.~G. Ryskin.
\newblock {Exclusive LHC physics with heavy ions: SuperChic 3}.
\newblock \emph{Eur. Phys. J. C}, 79\penalty0 (1):\penalty0 39, 2019.
\newblock \doi{10.1140/epjc/s10052-018-6530-5}.

\bibitem[Klusek-Gawenda et~al.(2016{\natexlab{a}})Klusek-Gawenda, Lebiedowicz,
  and Szczurek]{Klusek-Gawenda:2016euz}
Mariola Klusek-Gawenda, Piotr Lebiedowicz, and Antoni Szczurek.
\newblock {Light-by-light scattering in ultraperipheral Pb-Pb collisions at
  energies available at the CERN Large Hadron Collider}.
\newblock \emph{Phys. Rev. C}, 93\penalty0 (4):\penalty0 044907,
  2016{\natexlab{a}}.
\newblock \doi{10.1103/PhysRevC.93.044907}.

\bibitem[Bern et~al.(2001)Bern, De~Freitas, Dixon, Ghinculov, and
  Wong]{Bern:2001dg}
Z.~Bern, A.~De~Freitas, Lance~J. Dixon, A.~Ghinculov, and H.~L. Wong.
\newblock {QCD and QED corrections to light-by-light scattering}.
\newblock \emph{JHEP}, 11:\penalty0 031, 2001.
\newblock \doi{10.1088/1126-6708/2001/11/031}.

\bibitem[Klusek-Gawenda et~al.(2016{\natexlab{b}})Klusek-Gawenda, Sch{\"a}fer,
  and Szczurek]{Klusek-Gawenda:2016nuo}
Mariola Klusek-Gawenda, Wolfgang Sch{\"a}fer, and Antoni Szczurek.
\newblock {Two-gluon exchange contribution to elastic $\gamma \gamma \to \gamma
  \gamma$ scattering and production of two-photons in ultraperipheral
  ultrarelativistic heavy ion and proton-proton collisions}.
\newblock \emph{Phys. Lett. B}, 761:\penalty0 399--407, 2016{\natexlab{b}}.
\newblock \doi{10.1016/j.physletb.2016.08.059}.

\bibitem[Todesco and Wenninger(2017)]{Todesco:2017nnk}
E.~Todesco and J.~Wenninger.
\newblock {Large Hadron Collider momentum calibration and accuracy}.
\newblock \emph{Phys. Rev. Accel. Beams}, 20\penalty0 (8):\penalty0 081003,
  2017.
\newblock \doi{10.1103/PhysRevAccelBeams.20.081003}.

\bibitem[Zyla et~al.(2020)]{ParticleDataGroup:2020ssz}
P.~A. Zyla et~al.
\newblock {Review of Particle Physics}.
\newblock \emph{PTEP}, 2020\penalty0 (8):\penalty0 083C01, 2020.
\newblock \doi{10.1093/ptep/ptaa104}.

\bibitem[Ebert et~al.(2003)Ebert, Faustov, and Galkin]{Ebert:2003mu}
D.~Ebert, R.~N. Faustov, and V.~O. Galkin.
\newblock {Two photon decay rates of heavy quarkonia in the relativistic quark
  model}.
\newblock \emph{Mod. Phys. Lett. A}, 18:\penalty0 601--608, 2003.
\newblock \doi{10.1142/S021773230300971X}.

\bibitem[Gupta et~al.(1996)Gupta, Johnson, and Repko]{Gupta:1996ak}
Suraj~N. Gupta, James~M. Johnson, and Wayne~W. Repko.
\newblock {Relativistic two photon and two gluon decay rates of heavy
  quarkonia}.
\newblock \emph{Phys. Rev. D}, 54:\penalty0 2075--2080, 1996.
\newblock \doi{10.1103/PhysRevD.54.2075}.

\bibitem[Ackleh and Barnes(1992)]{Ackleh:1991dy}
E.~S. Ackleh and Ted Barnes.
\newblock {Two photon widths of singlet positronium and quarkonium with
  arbitrary total angular momentum}.
\newblock \emph{Phys. Rev. D}, 45:\penalty0 232--240, 1992.
\newblock \doi{10.1103/PhysRevD.45.232}.

\bibitem[Lyons et~al.(1988)Lyons, Gibaut, and Clifford]{Lyons:1988rp}
Louis Lyons, Duncan Gibaut, and Peter Clifford.
\newblock {How to Combine Correlated Estimates of a Single Physical Quantity}.
\newblock \emph{Nucl. Instrum. Meth. A}, 270:\penalty0 110, 1988.
\newblock \doi{10.1016/0168-9002(88)90018-6}.

\bibitem[Valassi(2003)]{Valassi:2003mu}
A.~Valassi.
\newblock {Combining correlated measurements of several different physical
  quantities}.
\newblock \emph{Nucl. Instrum. Meth. A}, 500:\penalty0 391, 2003.
\newblock \doi{10.1016/S0168-9002(03)00329-2}.

\bibitem[Nisius(2014)]{Nisius:2014wua}
Richard Nisius.
\newblock {On the combination of correlated estimates of a physics observable}.
\newblock \emph{Eur. Phys. J. C}, 74\penalty0 (8):\penalty0 3004, 2014.
\newblock \doi{10.1140/epjc/s10052-014-3004-2}.

\bibitem[Nisius(2020)]{Nisius:2020jmf}
Richard Nisius.
\newblock {BLUE: combining correlated estimates of physics observables within
  ROOT using the Best Linear Unbiased Estimate method}.
\newblock 2020.
\newblock \doi{10.1016/j.softx.2020.100468}.

\bibitem[Klein and Nystrand(2004)]{Starlight:2004}
S.~R. Klein and J.~Nystrand.
\newblock Photoproduction of quarkonium in proton-proton and nucleus-nucleus
  collisions.
\newblock \emph{Phys. Rev. Lett.}, 92:\penalty0 142003, 2004.
\newblock \doi{10.1103/PhysRevLett.92.142003}.

\bibitem[Klein et~al.(2017)Klein, Nystrand, Seger, Gorbunov, and
  Butterworth]{Klein:2016yzr}
Spencer~R. Klein, Joakim Nystrand, Janet Seger, Yuri Gorbunov, and Joey
  Butterworth.
\newblock {STARlight: A Monte Carlo simulation program for ultra-peripheral
  collisions of relativistic ions}.
\newblock \emph{Comput. Phys. Commun.}, 212:\penalty0 258--268, 2017.
\newblock \doi{10.1016/j.cpc.2016.10.016}.

\end{thebibliography}
\clearpage
%%%%%%%%%%%%%%%%%%%%%%%%%%%%%%%%%%%%%%%%%%%%%%%%%%%%%%%%%%%%%%%%%%%%%%%%%%
\appendix
\section{Systematic uncertainties in the input LbyL cross-section measurements}
\label{app:allUncs}
The \myprocess\ fiducial cross-sections measured at \cmenergyComb by the ATLAS and CMS Collaborations, as well as their uncertainties, are summarized in Table~\ref{tab:LbyLXS}. Similarly to the approach that is followed in combinations using the BLUE method, the category subtotal and total uncertainties in the table are evaluated as the sum in quadrature of the individual uncertainties. To obtain the impact of each source of uncertainty, the BLUE method takes into account their correlations. The method used by each input analysis to evaluate the individual uncertainties is briefly described below.

The CMS result has a larger data statistical uncertainty than the ATLAS result because the two experiments use data samples of different sizes.
An additional uncertainty of 10\% in the final cross-section in CMS is considered as statistical uncertainty, reflecting the finite size of the data sample at high acoplanarities
used for the absolute normalization of the central exclusive production (CEP) plus residual nonexclusive background.

In the ATLAS measurement, the uncertainty due to the choice of LbyL signal MC generator is estimated by using an alternative signal
MC sample. This affects the signal yield by 1\% which is taken as a systematic uncertainty, whereas the uncertainty due to the limited signal MC sample size is also considered and found to be 1\%.
In the CMS measurement, the difference in the fiducial $\PbPb\,(\mygg)\to \textrm{Pb}^{(\ast)}\textrm{+}\textrm{Pb}^{(\ast)}\,\ee$ cross-sections between the STARLIGHT and SuperChic MC generators amounts to 10\%.

The uncertainty in the integrated luminosity of the 2015+2018 data sample from ATLAS is 3.2\%.
It is derived from the calibration of the luminosity scale using horizontal--vertical beam-separation scans.

In the background-determination category, the uncertainty is estimated using a fully data-driven method in the ATLAS measurement.
The resulting uncertainty is dominated by the limited event yield in the dedicated control region, while the uncertainty also accounts for the extrapolation to the LbyL signal region.
The normalization of the CEP and exclusive \ee\ background in the LbyL signal region propagates into a 6\% uncertainty in the CMS cross-section measurement,
accounting for the finite size of MC samples.

The uncertainties in the photon reconstruction and identification efficiencies are estimated by parameterizing the scale factors as a function of the photon pseudorapidity, affecting the expected LbyL signal yield in the ATLAS measurement by 4\% (``photon reco. efficiency'') and 2\% (``photon PID efficiency'').
The main sources of uncertainty in the LbyL scattering and exclusive \ee\ production measurements from CMS are related to the single photon (9\%) and electron (2.5\%)
reconstruction and identification efficiencies. These uncertainties are doubled in the total cross-section, since CMS considers diphoton and dielectron final states.
The uncertainties related to the photon energy scale and resolution affect the expected signal yield by 1 and 2\%, respectively, in the ATLAS measurement.
The procedure implemented by the CMS Collaboration effectively includes bin migrations outside the fiducial \Et range due to the effects of photon energy scale and resolution.
However, the impact of bin migrations in the final diphoton cross-section is found to be below 1\%.
The uncertainty due to imperfect knowledge of the photon angular resolution results in a 2\% shift of the signal yield in the ATLAS measurement.
Systematic uncertainties in the trigger efficiency are dominated by the statistical uncertainty of each data set.
In total, the impact of the trigger efficiency uncertainty on the expected LbyL signal yields is 5 and 12\% in the ATLAS and CMS measurements, respectively. 

\begin{table*}[!thp]
  \begin{center}
    \caption{Measured fiducial cross-sections, uncertainty components and their
      magnitudes (relative to the individual measurements) for the ATLAS and CMS $\myprocess$ 
      measurements at \sqrtNN~=~\cmenergyComb. 
      The CMS measurement is marked with $^\dagger$ for its scaling by a correction factor to account for differences in the definition of phase space regions, as described in Section~\ref{sec:theory}.
      Uncertainties in the same category can
      be compared between experiments, as detailed in the text. The
      naming conventions follow those of the corresponding experiments. The category subtotal and total uncertainties are emphasized, and are evaluated as the sum in quadrature of the individual uncertainties.}
    \begin{adjustbox}{max width=1.0\textwidth}{}
      \begin{tabular}{l | l | r | l | r}
\hline
\hline
\T\B
 & \multicolumn{2}{c|}{ATLAS~\cite{Aad:2020cje}} &   \multicolumn{2}{c}{CMS~\cite{Sirunyan:2018fhl}}  \\ 
\hline\hline
Cross-section &  \multicolumn{2}{c|}{120~nb} & \multicolumn{2}{c}{91$^\dagger$~nb} \\
\hline\hline
Uncertainty category &\multicolumn{2}{c|}{Uncertainty [\%]} &\multicolumn{2}{c}{Uncertainty [\%]}  
\\\hline
Statistical& Data statistical & 14& Data statistical & 37 \\
&  &  & CEP and QED bkg. normalization & 10 \\
\hline
\hspace{+2mm}Category subtotal & \multicolumn{2}{r|}{\bf 14} & \multicolumn{2}{r}{\bf 38} \\
\hline
\hline
Theory modeling        
& Signal MC statistical& 1 &  &  \\
& Alternative signal MC& 1 &  &  \\
&  & & Derivation of $\sigmatheo(\ggee)$& 10\\
\hline
\hspace{+2mm}Category subtotal & \multicolumn{2}{r|}{\bf 1} & \multicolumn{2}{r}{\bf 10}  \\
\hline
\hline       
Integrated luminosity& &  3& & \\
\hline
\hspace{+2mm}Category subtotal & \multicolumn{2}{r|}{\bf 3} & \multicolumn{2}{r}{} \\  
\hline
\hline
Background determination
& Data-driven $\ggee$ method & 6& Size of simulated background samples & 6 \\
\hline
\hspace{+2mm}Category subtotal & \multicolumn{2}{r|}{\bf 6} & \multicolumn{2}{r}{\bf 6} \\
\hline
\hline
Photon reconstruction and identification& Photon reco. efficiency & 4&  &  \\
& Photon PID efficiency & 2&  &  \\
& Photon energy scale & 1&  &  \\
& Photon energy resolution & 2&  &  \\
& & & Photon reco.$\oplus$ID efficiency & 18 \\
\hline
\hspace{+2mm}Category subtotal & \multicolumn{2}{r|}{\bf 5} & \multicolumn{2}{r}{\bf 18} \\
\hline
\hline
Photon angular resolution
& Photon angular resolution & 2&  &  \\
\hline
\hspace{+2mm}Category subtotal & \multicolumn{2}{r|}{\bf 2} & \multicolumn{2}{r}{} \\
\hline
\hline
Electron reconstruction and identification
&  & &Electron reco.$\oplus$ID efficiency  & 5 \\
\hline
Category subtotal & \multicolumn{2}{r|}{\bf } & \multicolumn{2}{r}{\bf 5 } \\
\hline
\hline
Trigger
& Trigger efficiency & 5 & Trigger efficiency & 12 \\
\hline
\hspace{+2mm}Category subtotal & \multicolumn{2}{r|}{\bf 5} & \multicolumn{2}{r}{\bf 12 } \\
\hline
\hline 
Total uncertainty & \multicolumn{2}{r|}{\bf 18} & \multicolumn{2}{r}{\bf 46}\\
\hline
\hline
\end{tabular}
       \end{adjustbox}{} 
    \label{tab:LbyLXS}
  \end{center}
\end{table*}

 \section{Systematic uncertainties and correlation assumptions}
\label{app:systcat}
In order to average the LbyL cross-section measurements, the sources of uncertainty are grouped into categories. Assumptions are made about correlations between similar sources of uncertainty in the two input measurements. Also uncertainties associated with theoretical predictions are taken into account for extracting the average. The correlations between similar uncertainties in theoretical predictions are discussed in Section~\ref{sec:systcat_theo}.

\subsection{Systematic uncertainties from experimental measurements}

Throughout this note, individual uncertainties are taken as reported by the input analyses, regardless of the method used to determine them. Although the sources of systematic uncertainty and the procedures used to estimate their impact on the measured cross-section are partially different in the individual analyses, it is still possible to identify contributions that describe similar physical effects. These contributions are listed below; they are grouped together, and only the resulting categories are used in the extraction of the averaged value. Categories are treated as uncorrelated among each other. For each source of uncertainty, correlations between the two measurements are assumed to be positive. The uncertainties in each category are listed below, with the correlation assumptions across the two input measurements given in parentheses.

\noindent {\bf Statistical} (Correlation 0) \\ This statistical uncertainty arises from the limited size of the data sample. It is uncorrelated between ATLAS and CMS. In the CMS measurement, an additional uncertainty of 10\% in the final cross-section, reflecting the finite size of the data sample at high \Aco, is used for the absolute normalization of the CEP background, and is also considered as statistical rather than systematic uncertainty.

\noindent {\bf Integrated luminosity} (Correlation 0)\\ This uncertainty originates from the systematic uncertainty in the integrated luminosity, as determined by the individual experiments using the methods described in Refs.~\cite{Sirunyan:2018fhl,Aad:2020cje}. It affects the normalization of signal yields in the ATLAS measurement, whereas since the ratio of cross-sections of the LbyL scattering over the exclusive $e^+e^-$ processes is measured by CMS, the uncertainty related to integrated luminosity is eliminated. The correlation coefficient between the integrated-luminosity uncertainty in ATLAS and CMS is thus 0. 

\noindent {\bf Background determination} (Correlation 0)\\
The uncertainty in the determination of the exclusive $e^+e^-$ plus any residual (nonacoplanar) background accounts for the finite size of the data (MC) samples in the ATLAS (CMS) measurement.
Therefore, the associated uncertainties are considered uncorrelated between ATLAS and CMS, despite the similar production modes.
The averaged \sigmameas result does not depend significantly on this correlation assumption.

\noindent {\bf Detector modeling} \\ This category is split into components and includes the uncertainty in the modeling of photons, electrons, and trigger. In the analyses, the uncertainties related to the reconstruction, identification, calibration (i.e., in terms of energy/angular scale and resolution), and trigger are propagated through variations of the procedures used to correct for effects from the detector response (``scale factors'').

\begin{itemize}
\item {\it Photon reconstruction and identification} (Correlation 0.5): Systematic uncertainties associated with the photon reconstruction and identification efficiency, photon energy scale and resolution are considered partially correlated between the ATLAS and CMS measurements: although independent data and MC samples are used, a similar methodology for the corrections is implemented. In the CMS measurement, no additional uncertainty in the photon energy scale and resolution is considered, since possible data to MC simulation differences are already included in the derivation of the reconstruction and identification scale factors.  To ensure the robustness of the results against the correlation assumptions for this large uncertainty, the extraction of the averaged value is performed with alternative correlation values.
\item {\it Photon angular resolution} (Correlation 0): In the ATLAS measurement, the data to MC simulation difference in the electron cluster $\phi$ resolution is applied as an extra correction to photons, resulting in a shift of the LbyL signal yield, which is taken as an uncorrelated systematic uncertainty between the two measurements.
\item {\it Electron reconstruction and identification} (Correlation 0): In the CMS measurement, the electron modeling uncertainty includes components associated with the electron reconstruction and identification efficiency. The incorporation of exclusivity efficiencies is irrelevant, because they cancel out in the ratio of cross sections of the LbyL scattering over the exclusive \ee\ processes. This uncertainty is thus considered uncorrelated between ATLAS and CMS, since it is of relevance for the CMS measurement only.
\item {\it The level 1 and high-level triggers} (Correlation 0): For both measurements systematic uncertainties in the trigger efficiency  are dominated by the statistical uncertainty of each data set and are thus uncorrelated. We found out that the averaged \sigmameas result does not depend significantly on that correlation assumption.
\end{itemize}

 \subsection{Systematic uncertainties in theoretical predictions} \label{sec:systcat_theo}

The systematic uncertainties in the averaged $\sigmameas$ value include uncertainties in theoretical predictions, and the correlation assumptions are explained below. \\

\noindent {\bf Theory modeling}\\ This category contains the uncertainties in the modeling of the simulated LbyL process, as well as contributions from the modeling of other processes because the uncertainties are closely related as explained in the following. These include uncertainties statistical and systematic in nature, and are summed in quadrature in each input measurement.

\begin{itemize}
\item {\it Simulation statistical} (Correlation 0): This statistical uncertainty comes from the limited size of simulated event samples. It is uncorrelated between ATLAS and CMS and has a negligible impact on the averaged \sigmameas.
\item {\it Simulation systematic} (Correlation 1): In the ATLAS measurement, an alternative LbyL signal sample was generated using calculations from Ref.~\cite{Klusek-Gawenda:2016euz}, with the difference between the signal predictions driven by the non-hadronic overlap condition of the Pb ions. In the CMS measurement, the fiducial LbyL cross section is obtained from the theoretical prediction of $\sigmatheo(\ggee)$ from the STARLIGHT~v2.76~\cite{Starlight:2004,Klein:2016yzr} generator, where the uncertainty is derived from SuperChic to compute the non-hadronic overlap condition in the MC simulation. Given the difference between the nominal and alternative predictions lies in the implementation of the non-hadronic overlap condition of the Pb ions, we considered this source of uncertainty as fully correlated between the two measurements. 
\end{itemize}

The contribution from each uncertainty category to the total uncertainty in the averaged cross-section measurement is given in Table~\ref{tab:Res}.

\begin{table*}[!htbp]
  \caption{Contribution from each uncertainty category in relative and absolute terms to the averaged
    LbyL cross-section (\sigmameas) uncertainty at \cmenergyComb.
    Correlations of uncertainties between the two input measurements are presented in~\ref{app:allUncs}.
    The total systematic uncertainties ``(excl. lumi.)'' and ``(excl. theo.)'' are computed by considering
    all the individual sources of systematic uncertainty excluding the uncertainty due to integrated luminosity and theory modeling, respectively.
    The total uncertainty includes the statistical uncertainty.
}
  \begin{center}
  \begin{tabular}{l | r | r}
\hline
 \hline
 \multicolumn{3}{c}{ $\sigmameas$, $\sqrtNN~=~\cmenergyComb$ } \\
\hline
\hline
{\bf Averaged cross-section} & \multicolumn{2}{c}{\finalResultCentVal} \\
\hline
\hline
\multirow{2}{*}{Uncertainty category} & \multicolumn{2}{c}{Uncertainty} \\\cline{2-3}
 &  ~~[\%] &  [nb] \\
\hline
Statistical & 13 & 15\\
\hline
Integrated luminosity & 3 & 3\\
Background determination & 5 & 6\\
Photon reconstruction and identification & 6 & 7\\
Photon angular resolution& 1 & 2\\
Electron reconstruction and identification& $<1$ & 1\\ 
Trigger & 5 & 5\\
Theory modeling &3 & 3\\
\hline
Total syst.\ unc.\ (excl.\ lumi.) & 9 & 11\\
Total syst.\ unc.\ (excl.\ theo.) & 9 & 11\\
Total syst.\ unc.\  & 10 & 12\\
\hline\hline
{\bf Total uncertainty} & \finalResultRelPrecision & \finalResultAbsPrecision\\
\hline
 \hline
  \end{tabular}
  \end{center}
\label{tab:Res}
\end{table*}

\subsection{Stability tests}
\label{sec:xscombtests}

The stability of the averaged result to variations in the correlation assumptions is checked for the dominant uncertainty contributions.
The correlation values are varied for the categories of background determination, photon reconstruction, and trigger. More specifically, the categories of background determination and trigger are independently varied from their default values of uncorrelated to half correlated and half anti-correlated and the more extreme variation of fully correlated.
For the category of photon reconstruction, the correlation between ATLAS and CMS is varied from its default value of 50\% correlated to fully correlated and uncorrelated and the more extreme variation of half anticorrelated.

Because of the scheme that is used for the correlations, stability tests are also performed for the uncertainty associated with the theory modeling.
The uncertainties in the theory modeling category are varied from their default value of fully correlated to half correlated and to the more extreme tests of uncorrelated and half anti-correlated.

\begin{figure*}[!h!tbp]
  \begin{center}
    \includegraphics[width=1.0\textwidth]{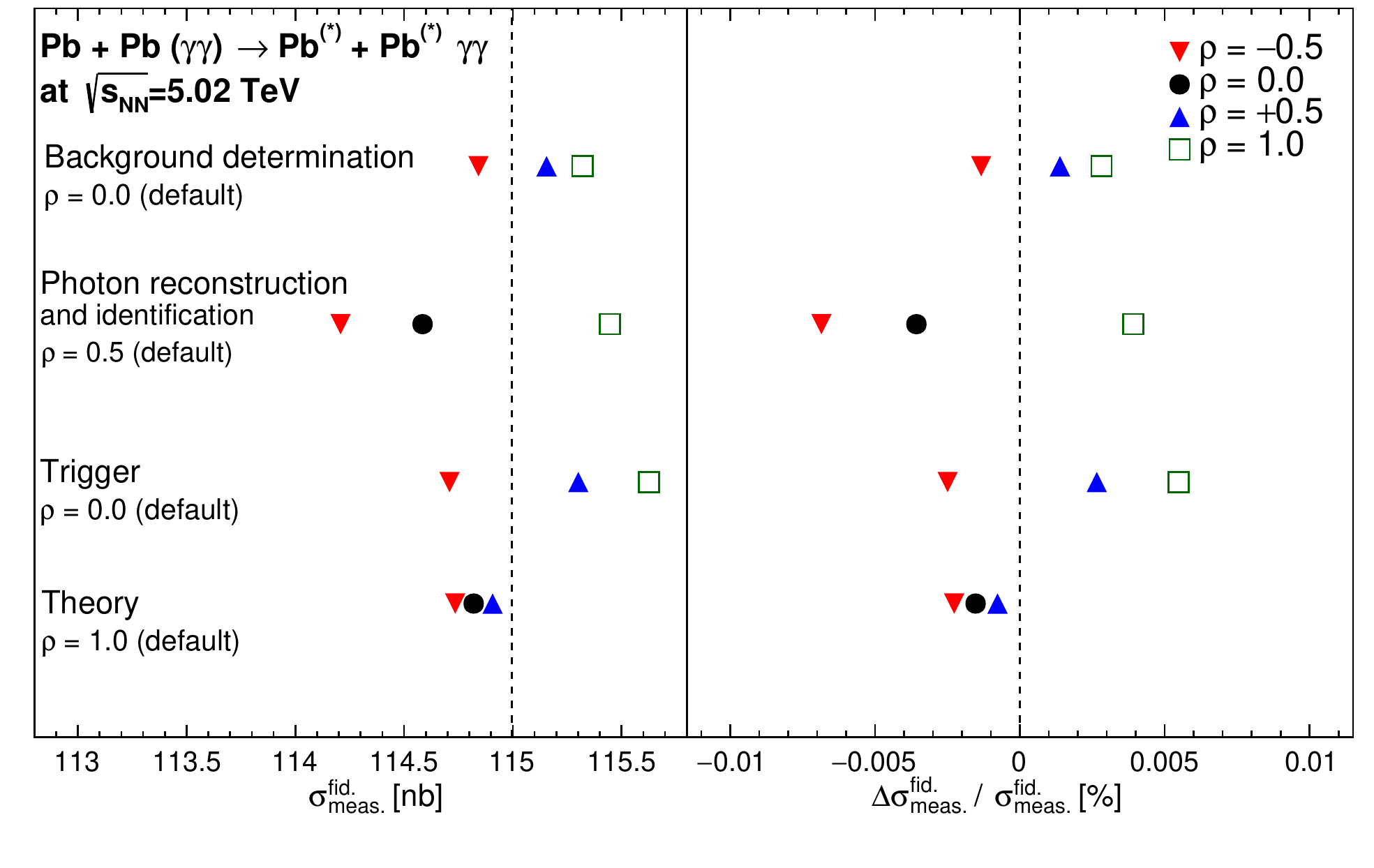}
    \caption{(left panel)~Results of the stability tests demonstrating impact of  variations of
    the correlation assumptions in different uncertainty
    categories on the averaged cross-section are shown. (right panel)~The corresponding relative shifts (with $\Delta$ = varied $-$ nominal) in the central value, \sigmafid, and in its uncertainty,
     $\Delta(\sigmafid)/(\sigmafid)$, are shown.}
    \label{fig:vtbstabilities}
  \end{center}
\end{figure*}

Figure~\ref{fig:vtbstabilities} summarizes the results of these stability tests, where the correlations between the two input measurements are varied.
For all variations, the relative changes in the central value of the averaged \sigmameas are insignificant relative to the total uncertainty of \finalResultRelPrecision.
These tests show that the averaged result is robust and does not critically depend on any of the correlation assumptions.

\end{document}